\documentclass[pra,aps,epsfig,graphics,twocolumn,twoside]{revtex4}
\usepackage{amsfonts}

\usepackage[dvips]{graphicx}
\usepackage[T1]{fontenc}

\usepackage[latin2]{inputenc}
\usepackage{amsmath}
\usepackage{mathtools}
\usepackage{latexsym}
\usepackage{bbm}

\usepackage{graphicx}
\usepackage{epsfig}
\usepackage{color}

\def\dt{{\rm d}\,}

\def\duzomniejsze{<\kern-.7mm<}
\def\duzowieksze{>\kern-.7mm>}

\def\textbf#1{{\bf #1}}
\def\beq{\begin{equation}}
\def\eeq{\end{equation}}
\def\be{\begin{equation}}
\def\ee{\end{equation}}
\def\ben{\begin{eqnarray}}
\def\een{\end{eqnarray}}
\def\beqa{\begin{eqnarray}}
\def\eeqa{\end{eqnarray}}
\def\eea{\end{array}}
\def\bea{\begin{array}}
\def\bem{\begin{matrix}}
\def\eem{\end{matrix}}
\newcommand{\bei}{\begin{itemize}}
\newcommand{\eei}{\end{itemize}}
\newcommand{\bee}{\begin{enumerate}}
\newcommand{\eee}{\end{enumerate}}

\def\1{\openone}

\def\>{\rangle}
\def\<{\langle}

\def\ot{\otimes}

\def\dt#1{{{\kern -.0mm\rm d}}#1\,}
\def\squareforqed{\hbox{\rlap{$\sqcap$}$\sqcup$}}
\def\qed{\ifmmode\squareforqed\else{\unskip\nobreak\hfil
\penalty50\hskip1em\null\nobreak\hfil\squareforqed
\parfillskip=0pt\finalhyphendemerits=0\endgraf}\fi}

\newtheorem{lemma}{Lemma}

\newtheorem{theorem}[lemma]{Theorem}
\newtheorem{proposition}[lemma]{Proposition}
\newtheorem{definition}[lemma]{Definition}

\newtheorem{fact}[lemma]{Fact}

\def\bep{\begin{proposition}}
\def\eep{\end{proposition}}
\def\bel{\begin{lemma}}
\def\eel{\end{lemma}}

\def\bet{\begin{theorem}}
\def\eet{\end{theorem}}
\def\bed{\begin{definition}}
\def\eed{\end{definition}}
\def\bef{\begin{fact}}
\def\eef{\end{fact}}

\begin{document}

\title{Simple scheme for encoding and decoding a qubit in unknown state for various topological codes}

\author{Justyna \L{}odyga$^{1}$, Pawe\l{} Mazurek$^{2,3}$, Andrzej Grudka$^{1}$, Micha{\l} Horodecki$^{2,3}$}

\affiliation{$^1$Faculty of Physics, Adam Mickiewicz University, 61-614 Pozna\'{n}, Poland\\
$^2$Institute for Theoretical Physics and Astrophysics,
University of Gda{\'n}sk, 80-952 Gda{\'n}sk, Poland\\
$^3$National Quantum Information Centre of Gda\'{n}sk, 81-824 Sopot, Poland
}

\begin{abstract}
We present a scheme for encoding and decoding an unknown state for CSS codes, based on syndrome measurements. We illustrate our method by means of Kitaev toric code, defected-lattice code, topological subsystem code and 3D Haah code. The protocol is local whenever in a given code the crossings between the logical operators consist of next neighbour pairs, which holds for the above codes. For subsystem code we also present scheme in a noisy case, where we allow for bit and phase-flip errors on qubits as well as state preparation and syndrome measurement errors. Similar scheme can be built for two other codes. We show that the fidelity of the protected qubit in the noisy scenario in a large code size limit is of $1-\mathcal{O}(p)$, where $p$ is a probability of error on a single qubit per time step. Regarding Haah code we provide noiseless scheme, leaving the noisy case as an open problem. 
\end{abstract}

\maketitle

\section{Introduction}
Interaction of a quantum system with environment leads to decoherence of its state. Protecting quantum state from decoherence can be regarded as the simplest quantum computation protocol. The so called threshold theorem \cite{Aharonov98} states that every quantum computation can be realised with arbitrary precision provided the error probability is below some threshold value, with polylogarithmic overhead in space and time. The assumption here is that the noise is local in a sense that error correlations decay exponentially both in space and time. Ref. \cite{Alicki13} presents a recent review of the Hamiltonian open system description of fault tolerant schemes.

Entanglement distribution in noisy quantum networks would allow for realisation of quantum cryptography tasks \cite{Bennett84,Ekert91} and distributed quantum computation \cite{Grover97}. Quantum repeaters idea \cite{Dur99,repeatersPRL} was one of the first approaches to entanglement distribution. It relies on application of entanglement purification protocol and entanglement swapping. The latter requires storage of quantum states in every node, with size of quantum memory scaling logarithmically with size of the network. Recently, a new idea for entanglement distribution that does not require local quantum memory was proposed. It is based on the isomorphism between storing quantum information in $D$-dimensional quantum network and establishing quantum communication in network of dimension $D+1$ \cite{perseguers-2008-78}. Namely, possibility of encoding, storing and decoding an unknown state 
of a physical qubit into $D$-dimensional network, implies possibility of communication over long distances using $D+1$-dimensional 
network. Actually, the task of protecting qubit is of independent interest on its own.

Stabilizer formalism \cite{Gottesman97}, analogical to construction of classical binary linear codes, offers a framework for description of many codes granting protection in the sense of threshold theorem.  Quantum information is stored here in the \textit{codespace} of total Hilbert space \footnote{Actually, in presence of errors, the system 
does not stay in the subspace, but rather in subsystem defined by so called {\it cosets}, 
which are subspaces obtained from the original subspace by applying errors; cf. \cite{BombinCHM2009-self-cor}.}. This logical  subspace is spanned by states that remain invariant under action of operators belonging to a stabilizer group that defines the code. In topological stabilizer codes, stabilizer group can be generated by local operators, which implies that logical subspace is protected from the local noise. One of the example of topological stabilizer code is a surface Kitaev code \cite{kitaevft}, which uses topological properties (the fact that operators acting on the logical subspace form loops on the physical space that are uncontractible to a point) to provide protection whenever the qubit error rate 
is below some threshold value. For longer logical operator lines (i.e. for bigger code) one has lower logical error rate.  In \cite{kitajewpreskill-ftmem,Horsman12}, authors present a way of encoding an unknown state into the Kitaev topological code by gradually enlarging the system of code qubits in a series of local CNOT operations. As the logical state of the code is vulnerable to decoherence for small code dimensions, this procedure requires fast qubit initialisation and measurements. Elaborated methods for creating maximally entanglement pairs between distant nodes of a $3$-dimensional cluster state network were considered in \cite{raussendorf-2005-71,Raussendorf06,perseguers2009-fid}. All these methods rely on measurements and operations actively performed on the code structure. Recently, Dengis, K\"{o}nig, and Pastawski proposed a scheme for dissipative state encoding into a Kitaev code, assuming Markovianity of bath \cite{dengis14}.             

In \cite{grudka12} a simple, active single shot scheme (which does not require gradual enlarging) for state encoding into a planar Kitaev code was proposed and it was shown to be equivalent to state teleportation into a code via the entangled pair of virtual qubits existing within the total Hilbert space. The protocol leads to lower bound on storing fidelity in large code size limit: $1-\mathcal{O}(p)$, with $p$ being the probability of bit-flip, phase-flip error as well as preparation and syndrome measurement errors in single time step. This enlightens the use of Kitaev planar code as quantum memory able to store an unknown state. Additionally, it provides an analytical lower bound for fidelity of quantum communication in 3D by the aforementioned isomorphism \cite{perseguers-2008-78}.

There exists a large variety of code constructions for quantum memory, focusing on different objectives. Among others, Kitaev toric code can store two qubits (in contrast with one qubit in its planar version) while keeping high threshold value \cite{kitajewpreskill-ftmem}. Planar codes with holes are able to store multiple qubits and enable CNOT operations by braiding \cite{Raussendorf07,Bombin09}. Topological subsystem codes \cite{Bravyi12} aim at increasing the locality in stabilizer measurements, while Haah in \cite{Haah11} introduces a three-dimensional topological code with no logical operators forming strings in physical code space.

In this paper we provide simple single shot protocol for encoding and decoding  an unknown qubit state for CSS codes. In the case when logical operators $X_L$ and $Z_L$ cross at a single physical qubit, 
our procedure requires preparation of qubits in a product state, and similarly, the final measurement is performed in a product basis.  
For CSS codes with logical operators crossing on larger number of physical qubits, it relies on preparation/measurement of entangled pairs of qubits. This can be performed locally if qubits where logical operators cross are situated on adjacent vertices of the lattice (we illustrate it using Haah code \cite{Haah11}). 

Our general procedure works in the case of ideal preparation and measurements. However it can be adapted to the case of non-ideal preparation and measurements for several topological codes: Kitaev toric code, planar codes with holes \cite{Raussendorf07,Bombin09} and topological subsystem codes \cite{Bravyi12}. We present it on the example of topological subsystem code. We show that in a noisy preparation and measurement scenario the bound on the error for encoding/storage/decoding process, that uses our scheme, is of order of noise acting on a single physical qubit per time step, in a large code size limit. We assume that the noise is local and take into account bit-flip and phase-flip storage, preparation and syndrome measurement errors.

The article is organised as follows. In Section II we present a review of a stabilizer formalism and introduce the general intuition behind the encoding/decoding scheme for a subgroup of stabilizer codes, namely CSS codes. In Section III we proceed with detailed description of the noiseless procedure for Kitaev topological code on torus, planar code with holes and subsystem topological code. We also consider the code without logical string operators introduced by Haah. In Section IV we describe in details noisy scenario for subsystem topological code and derive analytical upper bound on the error for this code. We conclude in Section V.

\section{Encoding/decoding procedure for CSS codes without noise}

\begin{figure}
\includegraphics[scale=0.5]{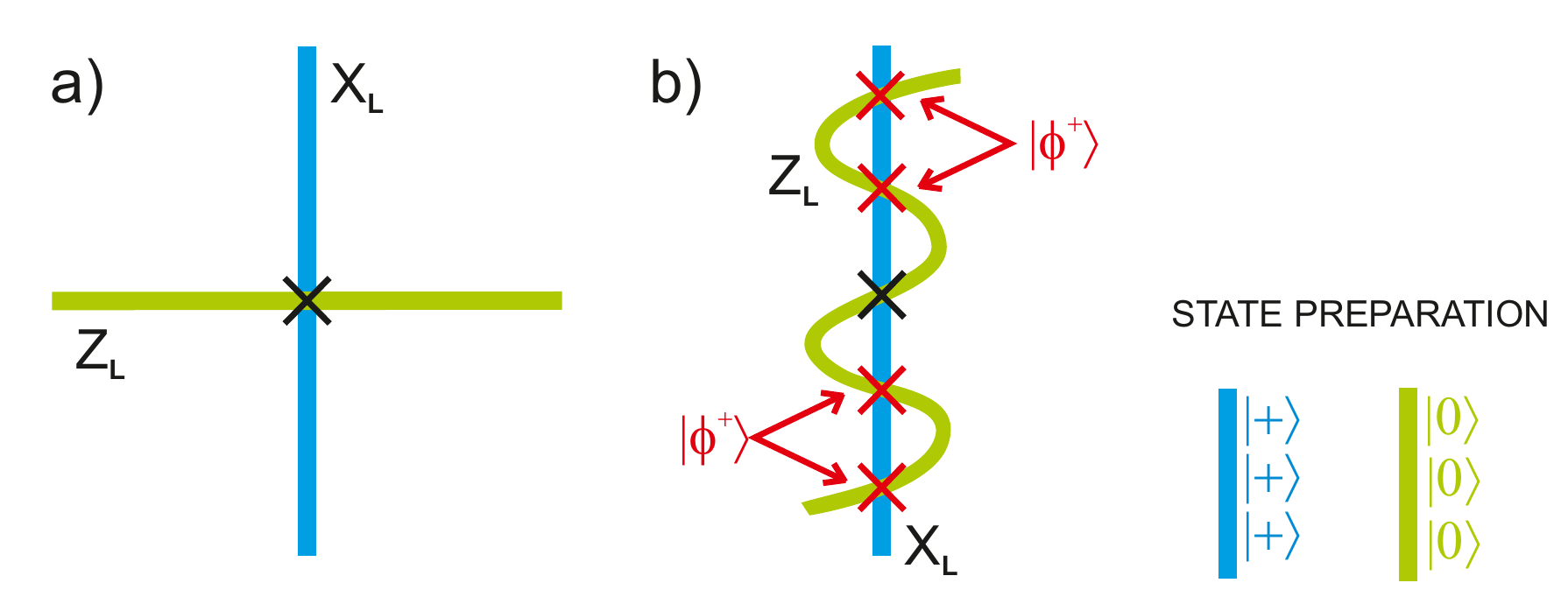}
\caption{\label{logical_operators}
\textbf{(a) Logical operators crossing at one physical qubit.} \textbf{(b) Logical operators crossing at many physical qubits.} Note that this is schematic picture. In reality the logical operators need not be the strings, and 
codes need not be planar. We put a qubit which we want to encode at one of crossing points (black cross). We prepare other qubits on which only one of logical operators $Z_{L}$ and $X_{L}$ acts nontrivially in states $|0\rangle$ and $|+\rangle$, respectively. In case of many crossing points we divide qubits on which both logical operators act nontrivially (red crosses) into pairs and prepare each pair in the maximally entangled state $\frac{1}{\sqrt{2}}(|0\rangle_{i,k}|0\rangle_{i,j}+|1\rangle_{i,k}|1\rangle_{i,j})$ where $(i,k)$ and $(i,j)$ label qubits composing each pair.}
  
\end{figure}

In this section we present the procedure of encoding and decoding qubit (or several qubits) for CSS codes, 
and show that it works perfectly when the noise is completely absent (i.e. preparation and measurements are ideal
and there are no storage errors).  
Before introducing the procedure, we briefly recapitulate the stabilizer formalism \cite{Gottesman97}
and the notion of CSS codes.

\textit{CSS codes.} The logical codespace $\mathcal{H}_{log}$, i.e. the subspace of the Hilbert space $\mathcal{H}_{system}=\otimes_{i}\mathcal{H}_{i}$ of $N$ qubits, with $i$-th qubit defined on Hilbert space $\mathcal{H}_{i}$ (dim$\mathcal{H}_{i}=2$), is spanned by eigenvectors with eigenvalue $1$ of the stabilizer group elements $\mathcal{S}$: $\{|\Psi\rangle:s|\Psi\rangle=|\Psi\rangle, \forall s\in\mathcal{S}\}$. Here $S$ is an abelian subgroup of Pauli group $P_{N}$ such that $-\mathcal{I}\not\in\mathcal{S}$, where $\mathcal{I}$ is an identity operator on $\mathcal{H}_{system}$, and $P_{N}$ is generated by $\{\sigma^{A}_{i}\}$, where bottom index indicates particular qubit $i=\left\{1,\dots,N\right\}$, upper index $A=\left\{X,Y,Z\right\}$ describes one of the Pauli matrices given by 
$X
=\begin{bmatrix}
0&1\\
1&0
\end{bmatrix}
$, $Y
=\begin{bmatrix}
0&-i\\
i&0
\end{bmatrix}
$ and  $Z
=\begin{bmatrix}
1&0\\
0&-1
\end{bmatrix},
$
and $\sigma^{A}_{i}=\mathcal{I}_{1}\otimes\dots\otimes\mathcal{I}_{i-1} \otimes A_{i}\otimes\mathcal{I}_{i+1}\dots\otimes\mathcal{I}_N$. $G(S)$, generator of $S$, can always be found to be the set of hermitian mutually commuting operators from the Pauli group $P_{N}$. Logical operators of the code are those operators from $P_{N}$ which commute with all operators from $G(S)$, but are not generated by them. Because $S$ is abelian, logical operators are defined \textit{modulo} $G(S)$. If we denote by $N-|G(S)|=D$, where $|G(S)|$ is the number of elements of G(S), then we can write dim$\mathcal{H}_{log}=2^D$ and $\mathcal{H}_{log}=\mathcal{H}_{L,1}\otimes\dots\otimes\mathcal{H}_{L,D}$. The set of operators commuting with $S$ is $\{Z_{L,1}, X_{L,1},\dots, Z_{L,D},X_{L,D}, S\}$. $Z_{L,i}, X_{L,i}$ create a pair of complementary observables (logical operators) acting on $\mathcal{H}_{L,i}$, i.e. $i$-th logical qubit subsystem of $\mathcal{H}_{log}$ (${\rm dim}\mathcal{H}_{L,i}$=2). They obey the following commutation and anticommutation relations: $Z_{L,i}X_{L,i}=-X_{L,i}Z_{L,i}$ and $[X_{L,i},Z_{L,j}]=0$ for $i\neq j$ as well as $[X_{L,i},X_{L,j}]=0$ and $[Z_{L,i},Z_{L,j}]=0$ for arbitrary $i$ and $j$.

We now consider one of $N$ physical qubits which is labelled by index $i\in[1,\dots,N]$. Without loss of generality we assume that 
it is in a $\alpha_i|0\rangle_{i}+\beta_i|1\rangle_{i}$ state (where $\{|0\rangle_{i},|1\rangle_{i}\}$ is the set of eigenvectors of $Z_{i}$), i.e. it is non entangled with other qubits. 
We define encoding $i$-th qubit
into the $i$-th logical qubit described on system  $\mathcal{H}_{L,i}$ as a process after which the $i$-th logical qubit is in the state  $|\Psi_{L_{i}}\rangle=\alpha_i|0\rangle_{L,i}+\beta_i|1\rangle_{L,i}$,
where $\{|0\rangle_{L,i},|1\rangle_{L,i}\}$ is the set of eigenvectors of a logical operator $Z_{L,i}$ on subsystem $\mathcal{H}_{L,i}$. We define decoding as reversed process.

Because the fidelity of quantum process depends only on the outcomes of measurements on two complementary sets of input states \cite{hofmann2004-fidelity}, to prove the correctness of encoding/decoding procedure it is enough to show that it performs a mapping between the eigenstates of $X_{i}$, $Z_{i}$ (acting on $\mathcal{H}_{i}$) and $X_{L,i}$, $Z_{L,i}$ (acting on $\mathcal{H}_{L,i}$), respectively.

The Calderbank-Shor-Steane (CSS) codes \cite{Shor95,Steane96,Calderbank96} are stabilizer codes that are characterised by the fact that their generator contains operators from $P_{N}$ that are only a tensor product of identity operators $I$ and either $X$ or $Z$ Pauli operators. This implies that logical operators are also of this form, therefore nontrivial elements of $Z_{L,i}$ involve only $Z$ operators, while nontrivial elements of $X_{L,i}$ are only given by $X$ Pauli operators. The anticommutation relation $Z_{L,i}X_{L,i}=-X_{L,i}Z_{L,i}$ is provided by the fact that $Z_{L,i}$ and $X_{L,i}$ cross on the odd number of physical qubits (act nontrivially on those qubits). A very useful property of CSS codes is that 
correction of phase errors is decoupled from correction of bit errors, so that one can consider them separately.

\textit{Logical operators crossing at a single qubit.} Let us now present a general idea for encoding a state into a subsystem $\mathcal{H}_{L,i}$ of a CSS stabilizer code, where $Z_{L,i}$ and $X_{L,i}$ cross at a single point. 

We select $\mathcal{H}_{i}$ arbitrarily and identify the corresponding vertex of the lattice with an intersection point of logical operators (Fig.\ref{logical_operators}(a)). Using the fact that in CSS codes $Z_{L,i}$ ($X_{L,i}$) is a tensor product of $Z$ ($X$) single qubit operators and identities, we make the parity of operators $Z_{L,i}$ ($X_{L,i}$) dependent only on the state of the $i$-th physical qubit. To this end we prepare all other qubits on which $Z_{L,i}$ ($X_{L,i}$) acts nontrivially (labelled here by $k$ ($l$)) in eigenstates associated with $+1$ eigenvalues of $Z_{k}$ ($X_{l}$). Since we assumed that $Z_{L,i}$ and $X_{L,i}$ operators cross at a single point, preparing procedures are independent. We will use the convention: $Z_{i}|0\rangle_{i}=+|0\rangle_{i}$, $Z_{i}|1\rangle_{i}=-|1\rangle_{i}$, $X_{i}|+\rangle_{i}=+|+\rangle_{i}$, $X_{i}|-\rangle_{i}=-|-\rangle_{i}$. Remaining qubits (i.e. those on which logical operators $Z_{L,i}$ and $X_{L,i}$ act trivially) are prepared in such a way that qubits on which $Z_{L,i}$ ($X_{L,i}$) acts nontrivially are surrounded by qubits in $|0\rangle$ ($|+\rangle$) states.

In order to drive a system state into a subspace $\mathcal{H}_{log}$, we 
measure stabilizer generators and join those of $Z$-type ($X$-type) that gave outcome $-1$ by chains of $X$ ($Z$) operators. Chains of $X$ ($Z$) that cross logical operators $Z_{L,i}$ ($X_{L,i}$) change their parity. However, as we can track the number of times it happens, we can revert this parity change by performing additional $X_{L,i}$ ($Z_{L,i}$) operation whenever this number is odd. Moreover, in specific cases of CSS codes considered in this paper it happens that the matching can always be performed in a way that does not affect the parity of logical operators and no additional corrections are needed at all. Therefore, the desired mapping $|0\rangle_{i}\rightarrow|0\rangle_{L,i}$, $|1\rangle_{i}\rightarrow|1\rangle_{L,i}$, $|+\rangle_{i}\rightarrow|+\rangle_{L,i}$, $|-\rangle_{i}\rightarrow|-\rangle_{L,i}$ is realised, where $|0\rangle_{L,i}$, $|1\rangle_{L,i}$, $|+\rangle_{L,i}$, $|-\rangle_{L,i}$ are eigenvectors of logical operators $Z_{L,i}$, $X_{L,i}$.

Decoding procedure of a logical qubit stored within $\mathcal{H}_{L,i}$ logical subspace of CSS code, with $Z_{L,i}$, $X_{L,i}$ logical operators crossing at a single physical qubit $i$, consists of performing single qubit measurement in $Z_{k}$ ($X_{l}$) basis on all the qubits where $Z_{L,i}$ ($X_{L,i}$) is nontrivially defined, except for the $i$-th physical qubit (Fig.\ref{logical_operators}(a)). From those measurements the parity of \textit{truncated} operator $Z_{T,i}$ ($X_{T,i}$) is calculated, where truncated operators are analogous to logical operators $Z_{L,i}$ and $X_{L,i}$, with the only difference that they act on $i$-th qubit trivially. If computed parity is odd, an operator $X_{i}$ ($Z_{i}$) is applied to the qubit defined on $\mathcal{H}_{i}$. This performs a demanded mapping $|0\rangle_{L,i}\rightarrow|0\rangle_{i}$, $|1\rangle_{L,i}\rightarrow|1\rangle_{i}$, $|+\rangle_{L,i}\rightarrow|+\rangle_{i}$, $|-\rangle_{L,i}\rightarrow|-\rangle_{i}$.  

\textit{Logical operators crossing on odd number of qubits.} Below we present an extension of the encoding/decoding procedure to topological stabilizer CSS codes where logical operators cross at more than one qubit.  This case is illustrated schematically in Fig.\ref{logical_operators}(b), where  
$Z_{L,i}$, $X_{L,i}$ operators act nontrivially on line of qubits. However, the following schemes are applicable to codes with 
arbitrary structure of logical operators. In addition, if logical operators $Z_{L,i}$ and $X_{L,i}$
cross at neighbouring qubits, then our encoding 
will be local. This is the case for Haah code \cite{Haah11}, where $Z_{L,i}$, $X_{L,i}$ are nontrivially defined on surfaces of $3$-dimensional rectangular lattice, with every vertex occupied by two qubits. To our best knowledge, no encoding scheme applicable to important class of Haah codes was proposed so far. 

The first step of the encoding procedure for codes with logical operators crossing at a single physical qubit was to make the parity of $Z_{L,i}$, $X_{L,i}$ dependent only on a state of physical qubit we want to encode. This was achieved by putting that qubit on the intersection of logical operators and preparing all other qubits $k$, on which $Z_{L,i}$ ($X_{L,i}$) acts nontrivially, in $|0\rangle_{k}$ ($|+\rangle_{k}$) states, stabilised by $Z_{k}$ ($X_{k}$). This was crucial for the state of $N$-qubit system to be an eigenstate of $\pm Z_{L,i}$ and $\pm X_{L,i}$, with signs dependent only on the sign of operators $\pm Z_{i}$ and $\pm X_{i}$, respectively, that were stabilising the $i$-th qubit.

Similarly we treat the case when logical operators cross at larger number of physical qubits. By $(i,k)$ we denote qubits on which at least one of logical operators  $Z_{L,i}$, $X_{L,i}$ acts nontrivially.  We choose a qubit to be encoded $(i,l)$ as one of the qubits at intersection points. As before, we prepare all physical qubits on which \textit{only one} logical operator acts nontrivially in the appropriate eigenstate of single qubit Pauli operators $Z_{i,k}$ (for $Z_{L,i}$) and $X_{i,k}$ (for $X_{L,i}$). Even number of qubits on which \textit{both} logical operators act nontrivially (not taking here into account the $(i,l)$ qubit) can always be divided into pairs consisting of qubits $(i,j_1)$ and $(i,j_2)$ that are prepared in eigenstates of $Z_{i,j_1}\otimes Z_{i,j_2}$ and $X_{i,j_1}\otimes X_{i,j_2}$ corresponding to eigenvalues $1$, i.e. maximally entangled states $\frac{1}{\sqrt{2}}(|0\rangle_{i,j_1}|0\rangle_{i,j_2}+|1\rangle_{i,j_1}|1\rangle_{i,j_2})$. Note that these operators commute. Such preparation scheme makes the parity of $Z_{L,i}$ ($X_{L,i}$) dependent only on the eigenvalue of $Z_{i,l}$ ($X_{i,l}$), as required. We drive the state of the system into $\mathcal{H}_{log}$ by performing a sequence of measurements and applying appropriate corrections (bit-flips and phase-flips operations) that either do not change the parity of logical operators (due to obeyed commutation relations) or change the parity (which can be fixed by applying additional logical operator to the code, as explained before).
        
Decoding procedure relies on measuring the parity of truncated operators $Z_{T,i}$ and $X_{T,i}$. In case of codes with logical operators crossing at one qubit $i$, the parity of truncated operators can be calculated from the measurements of single qubit Pauli operators, as there is no qubit $k$ enforcing anticommutation relation of $Z_{k}$ and $X_{k}$ measurements. Decoding procedure for codes with logical operators crossing at larger number of qubits relies on the same idea for solving the noncommutativity problem as the encoding one: we divide an even number of qubits on which both truncated operators act nontrivially, and perform commuting measurements of $Z_{i,j_1}\otimes Z_{i,j_2}$ and $X_{i,j_1}\otimes X_{i,j_2}$.
After the parity of truncated logical operators is calculated, the $(i,l)$ qubit is flipped by $X_{i,l}$ ($Z_{i,l}$) if the parity of $Z_{T,i}$ ($X_{T,i}$) is odd.

As said above, our encoding/decoding schemes rely on \textit{local} entanglement preparation and measurements for every code with intersection points of logical operators situated on adjacent vertices of the lattice, which is the case of the Haah code.

\section{Examples}

\subsection{Kitaev toric code}
\begin{figure}
\includegraphics[scale=0.5]{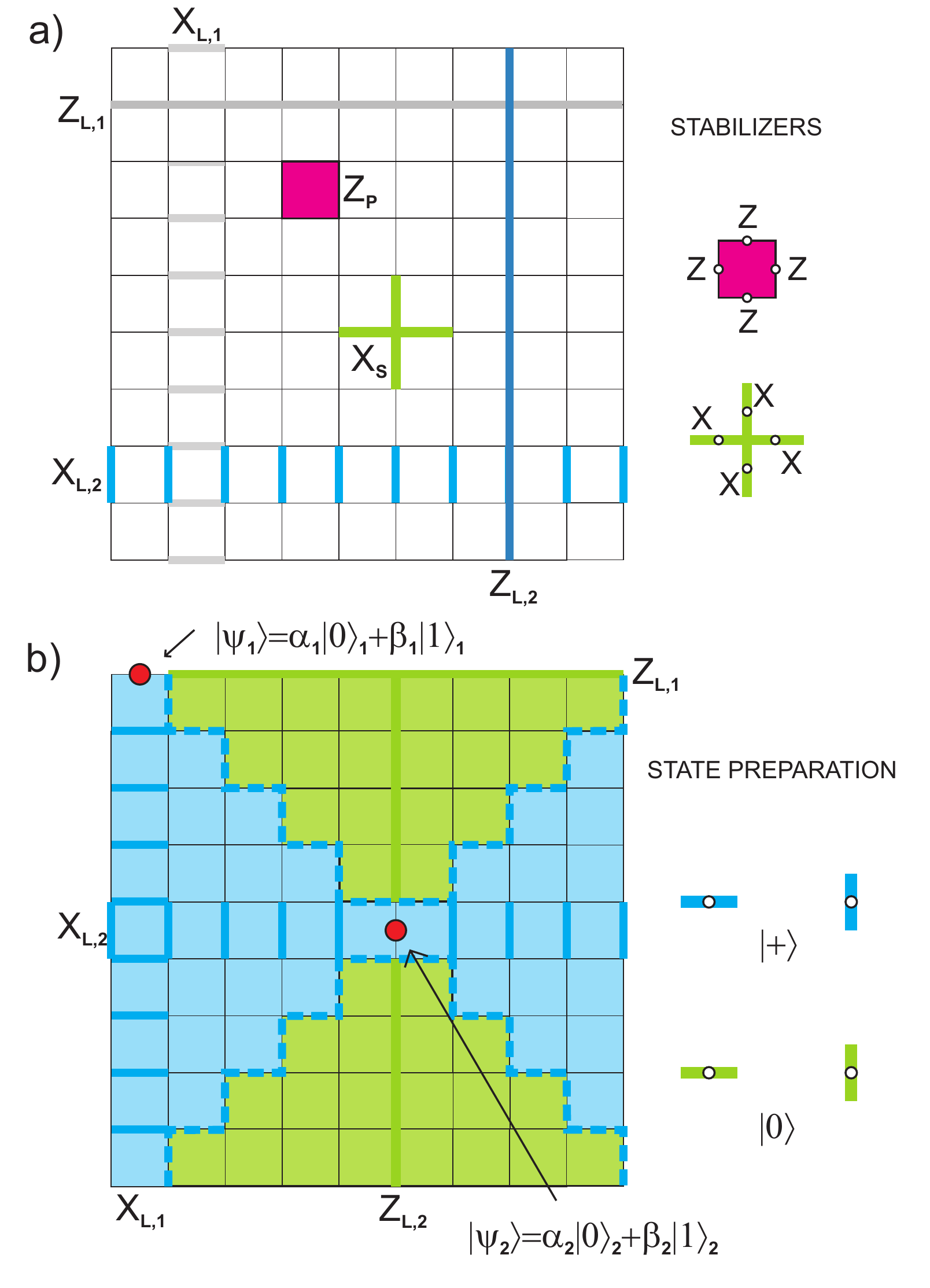}
\caption{\label{kitaev}
\textbf{(a) Kitaev toric code.} Each link is associated with a code qubit. The code is realized in the $L \times L$ square sheet on the torus, hence there are $2 L^2$ code qubits. Stabilizer generators of $X$-type are represented by stars (tensor product of $X$'s acting on four qubits building a particular star) and of $Z$-type by plaquettes (tensor product of $Z$'s acting on four qubits which merge to a particular plaquette). Due to periodic boundary conditions there are $2(L^2 -1)$ independent stabilizer generators. Thus, one is able to encode $2$ qubits. Exemplary logical operators associated with those qubits are marked in a picture in such a way that $X_{L,i}$ and $Z_{L,i}$ anticommute, where $i=1,2$. \textbf{(b) Preparation of a lattice.} Qubits to be encoded are marked by red dots. Qubits in green and blue regions are initialised in states $|0\rangle$ and $|+\rangle$, respectively.}
\end{figure}

Below we investigate a topological code discovered by Kitaev \cite{kitaevft} and developed in \cite{kitajewpreskill-ftmem}. 
In the particular case of its planar version, the encoding procedure was already presented in \cite{grudka12}. Using similar approach, we show here how to encode logical qubits in its toric architecture.

\textit{Code geometric structure.}  Qubits are situated on links of a 2D lattice with periodic boundary conditions (see Fig. \ref{kitaev}(a)). Stabilizer group $\mathcal{S}$ is generated by local four-qubit observables of two types -- the plaquette observables $Z_p$ and the star observables $X_s$:  
\be
X_s=\ot_{l\in s } \sigma^x_{l},\quad
Z_p=\ot_{l\in p } \sigma^z_{l}
\ee
Here $s$ stands for a star associated with a vertex and it denotes all links that touch the vertex, while $p$ stands for plaquette and it denotes all links that form the plaquette. Logical qubit operators are defined by lines of Pauli operators uncontractible to the point. There are two types of them: $X_{L,1}$, $Z_{L,1}$ and $X_{L,2}$, $Z_{L,2}$, as there are two logical qubits encoded. $X_{L,i}$ and $Z_{L,i}$ cross at one point, which provides correct anticommutation relation. Let us assume that a distance of the code is an odd number. As depicted in Fig. \ref{kitaev}(b) we draw a dashed line from top left corner of the lattice to the bottom right one, and the other line from top right corner to bottom left one. On the intersection points of two dashed lines (due to periodic boundary conditions, there are two such points) we insert two physical qubits which we want to encode. We choose a pair of logical operators $X_{L,i}$ and $Z_{L,i}$ that meet at a point occupied by each inserted qubit. The remaining qubits on which given operator $X_{L,i}$ ($Z_{L,i}$) acts nontrivially, are initialised in $|+\rangle$ ($|0\rangle$) states. We prepare the remaining qubits surrounding $X_{L,i}$ ($Z_{L,i}$) in $|+\rangle$ ($|0\rangle$) states; qubits on the boundary dashed line are prepared in $|+\rangle$ states, see Fig. \ref{kitaev}(b). This way we lock $X_{L,i}$ and $Z_{L,i}$ operators within $|+\rangle$ and $|0\rangle$ states, respectively. This ends the stage of lattice preparation for the encoding procedure.

\textit{Encoding procedure.}  Now we are ready to encode two physical qubits into the code. Firstly, local stabilizer measurements are performed. Since $X_s$ and $Z_p$ commute, we can consider the protection against phase-flip errors (detected by $X_s$ measurements) and bit-flip errors (detected by $Z_p$ measurements), separately. We measure $X_s$ ($Z_p$) stabilizers associated with each vertex (plaquette) of the lattice and store defects, i.e nontrivial outcomes, in the vertices of the original lattice (virtual lattice, where plaquettes are replaced with stars). In case of noiseless syndrome measurement and state preparation, it is sufficient to measure only such $X_s$ ($Z_p$) stabilizers that include at least one green (blue) qubit since the outcomes of remaining syndrome measurements are already known. Afterwards, we remove $X$ ($Z$)-defects by joining them by chains of Pauli operators $Z$ ($X$) so that we do not cross lines of logical operators $X_{L,i}$ ($Z_{L,i}$) that anticommute with the chains. It is always possible to find such paths due to periodic boundary conditions. This ensures that the parity of logical operators $X_{L,i}$ ($Z_{L,i}$) stays intact throughout our encoding procedure. Since the initial parity is equal to the phase (bit) value of selected red qubit, we are guaranteed that states of the physical qubits are directly encoded as logical qubits ($\alpha_{i}|0\rangle_i+\beta_{i}|1\rangle_i \rightarrow \alpha_{i}|0\rangle_{L,i}+\beta_{i}|1\rangle_{L,i} $). It can be shown that similarly one can encode three physical qubits into three dimensional version of a considered code, introduced in \cite{kitajewpreskill-ftmem}. Different codes with periodic boundary conditions can be addressed in this manner as well.
We note, that a similar idea of encoding into toric code (although using a different approach of dissipative encoding) was put forward independently in \cite{dengis14}.

\textit{Decoding procedure.} For each logical qubit the same following procedure is applied. We measure the qubits arranged along the logical operator $X_{L,i}$ ($Z_{L,i}$) in $|+\rangle$, $|-\rangle$ ($|0\rangle$, $|1\rangle$) basis, with exception of the red qubit. Next, the parity of outcomes is computed and if it is odd, phase (bit)-flip is applied to the red qubit. To prove that this procedure correctly decode a qubit let us focus on decoding the first logical qubit when the code is in the state $|0\rangle_{L,1}|\psi\rangle_{L,2}$. Then the parity of line where logical operator $Z_{L,1}$ is defined is even. Hence, measuring all the qubits belonging to that line except for the red qubit $1$ we obtain that the parity is equal to the bit of the red qubit. Since we wish to decode $|0\rangle_{L,1}$ into $|0\rangle$ state we have to flip the red qubit, when obtained parity is odd. The explanation for $|1\rangle_{L,1}$ and $|\pm\rangle_{L,1}$ encoded states is analogous.\\

\subsection{Planar code with holes}
\begin{figure}
\includegraphics[scale=0.5]{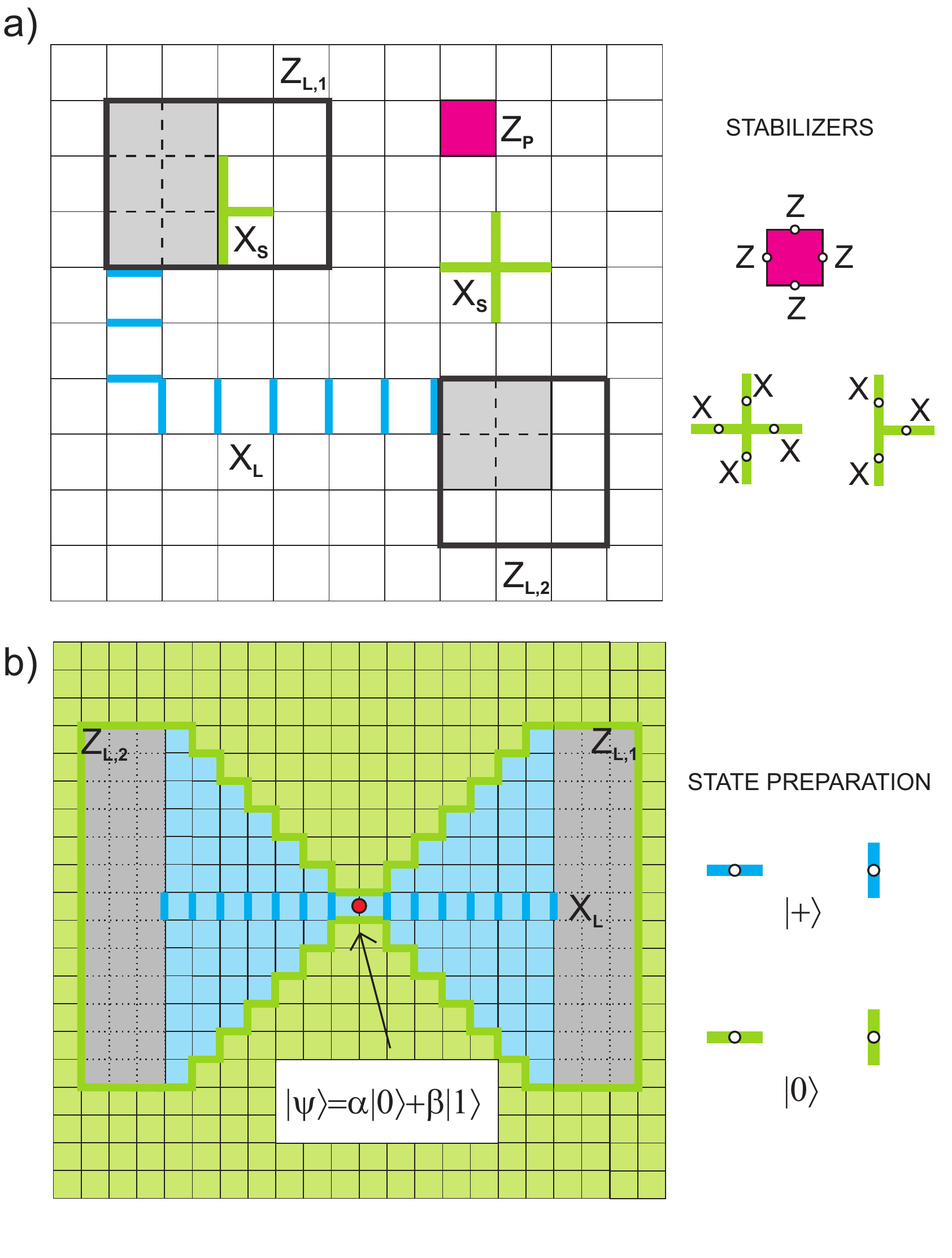}
\caption{\label{holes}
\textbf{(a) Planar code with holes.} Links denote code qubits. Stars denote stabilizer generators of $X$-type and plaquettes $Z$-type stabilizer generators. Topology is modified by removing some qubits and as a result creating a hole. Star operators centred on the edge of the hole are changed into $3$-weight star operators. Each hole is associated with logical qubit (where $Z_{L,i}$ is any closed loop around the hole, and $X_{L,i}$ goes from the hole and ends at the boundary). We are using these two holes to define one logical qubit with $Z_{L,1}$ or $Z_{L,2}$ as $Z_{L}$ operator and $X_{L,1}X_{L,2}$ (string of X's joining two holes) as $X_{L}$. \textbf{(b) Preparation of a lattice.} Qubit to be encoded is marked by red dot. Qubits in green and blue regions are initialised in states $|0\rangle$ and $|+\rangle$, respectively.}
\end{figure}

Another example of the code suited to the proposed encoding/decoding scheme is the one proposed in \cite{Raussendorf07,Bombin09}, enabling CNOT logical operations by braiding.

\textit{Code geometric structure.} Stabilizer group is generated by $4$-weight star and plaquette operators acting on a planar rectangular lattice as depicted in Fig. \ref{holes}(a). Qubits are placed on the links of the lattice. The code is deformed in such a way that in two regions stabilizers are not measured, i.e. holes are created, and these $4$-weight star operators which interfere into the 'empty' region are turned into $3$-weight star operators. Logical operators are such that $X_{L}$ is a chain of $X$ Pauli operators joining the holes and $Z_{L}$ is $Z_{L,1}$ or $Z_{L,2}$, where $Z_{L,i}$ is any closed loop of $Z$ Pauli operators around $i$-th hole. Let us encode one qubit into this code starting from a lattice arranged in the following way. Firstly, we have to choose the shape of logical operators $Z_{L,1}$ and $Z_{L,2}$ which will prevent from occurrence of infinitely many short nontrivial loops, which becomes relevant in a noisy scenario where such loops are dangerous. Hence, lines of logical operators have to be separated as much as possible and should meet at one point where we place a qubit to be encoded (the red one), see Fig. \ref{holes}(b). Operator $X_{L}$ joins two holes together and goes through a red qubit. Similarly to Kitaev toric code, here, lattice is divided into regions of qubits in $|+\rangle$ states (marked in blue, which belong to $X_{L}$ operator and its neighbourhood confined by lines of logical $Z_{L,1}$ or $Z_{L,2}$ logical operators) and qubits in $|0\rangle$ states (remaining qubits outside the holes, marked in green).

\textit{Encoding procedure.} If we want to encode an unknown state $|\Psi\rangle=\alpha|0\rangle+\beta|1\rangle$ of a physical qubit into a logical qubit, we have to make sure that the parity measured by logical operators depends on this state (whichever $Z_{L,i}$ we choose as our logical phase flip operator). In order to drive the system into the codespace we measure all $Z_{p}$ and $X_{s}$ stabilizers (because we are working in a noiseless scenario we may measure only such $Z_{p}$ ($X_{s}$) which touch at least one qubit in a $|+\rangle$ ($|0\rangle$)) state and correct syndromes by joining them by chains of $X$ ($Z$) Pauli operators. This can always be done without crossing the line of the logical operator anticommuting with the chain thus without changing its parity. From this we see that the state of logical qubit depends only on $|\Psi\rangle$, what fulfils the requirement of successful encoding.

\textit{Decoding procedure.} Similarly to the example described above, all qubits lying along the logical operator $X_{L}$ ($Z_{L}$) except for red one are measured in $|+\rangle$, $|-\rangle$ ($|0\rangle$, $|1\rangle$) basis. If parity calculated from the results is odd we have to apply phase (bit)-flip to the red qubit.\\

\subsection{Bravyi subsystem code}
\begin{figure}
\includegraphics[scale=0.5]{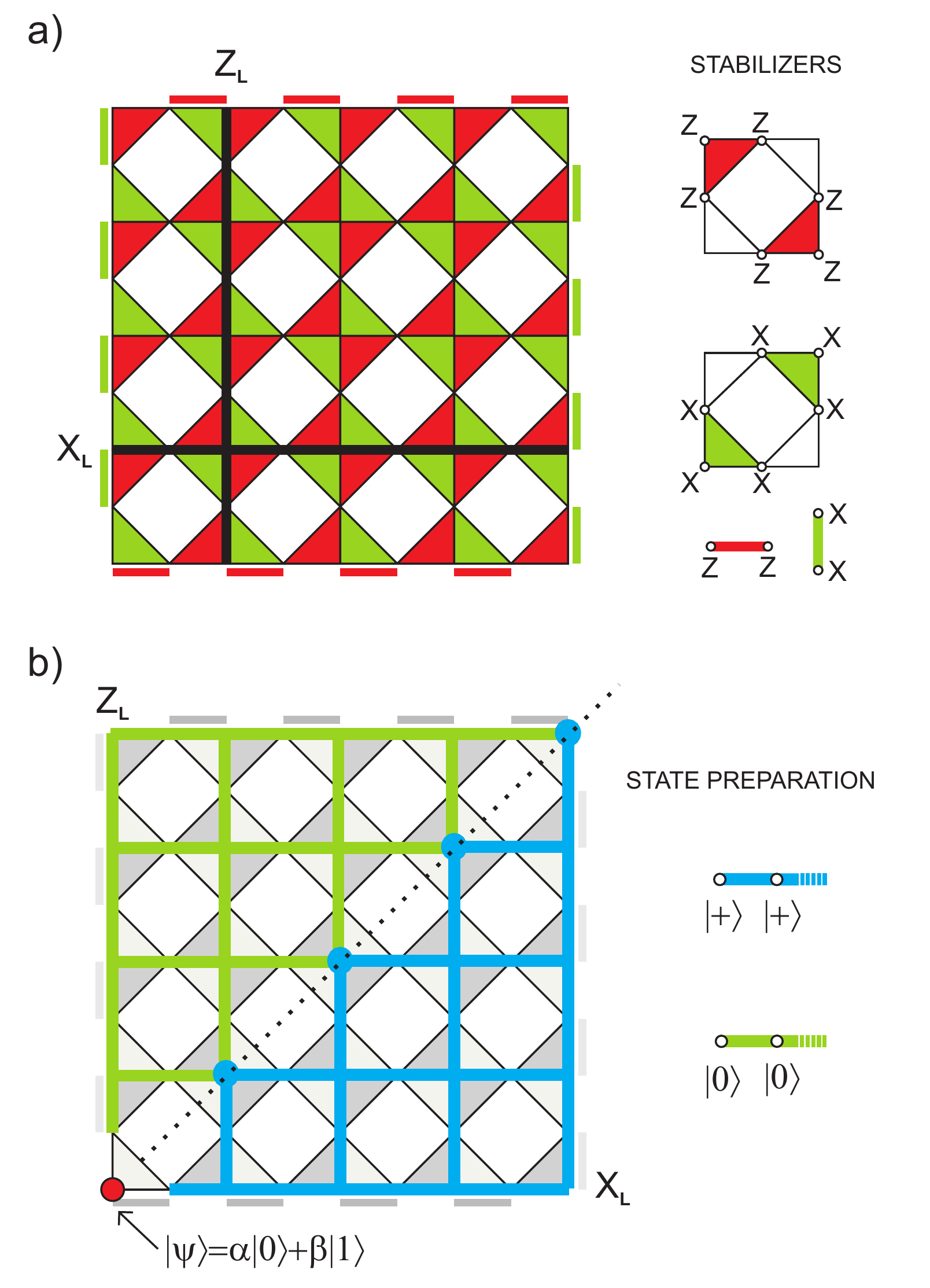}
\caption{\label{bravyi}
\textbf{(a) Subsystem surface code with qubits on the vertices.} There are two classes of check operators, i.e. $2$-weight operators at the boundaries (Z-type ones situated horizontally and X-type located vertically) and $6$-weight check operators made by two $3$-weight triangle operators merged together (and located diagonally in each cell for Z-type stabilizer generators and anti-diagonally for X-type). Logical operators (black lines) commute with every stabilizer and anticommute with each other. \textbf{(b) Preparation of a lattice.} Qubit to be encoded is marked by red dot. Qubits in green and blue regions are initialised in states $|0\rangle$ and $|+\rangle$, respectively.}
\end{figure}

\begin{figure}
\includegraphics[scale=0.5]{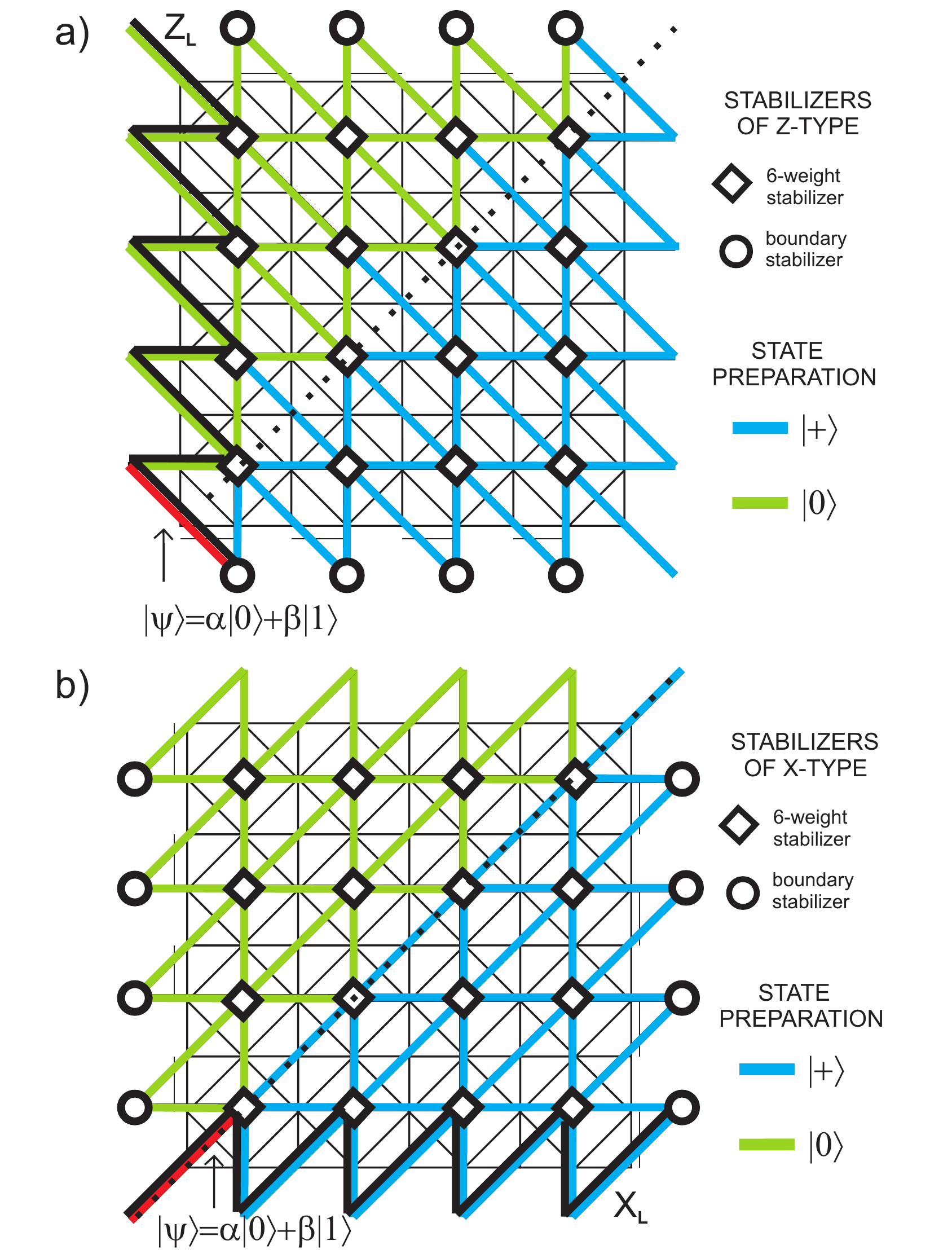}
\caption{\label{bravyi_virtual}
\textbf{(a) The virtual lattice of a subsystem surface code for bit-flip error correction.} Every diamond (denoting 6-weight Z-type stabilizer) is placed in the centre of every cell of the original lattice. From each diamond arises $6$ lines representing code qubits. 2-weight boundary check operators (associated with circles) close the lattice from up and down. Black thick line represents $Z_{L}$ logical operator. The code is prepared in the same manner as in Fig. \ref{bravyi}(b). \textbf{(b) The virtual lattice of a subsystem surface code for phase flip-error correction.} Preparation of the code is the same as in Fig. \ref{bravyi_virtual}(a) but with lattice rotated by $90$ degrees. Diamonds and circles depict stabilizers of X-type. Black thick line represents $X_{L}$ logical operator.}
\end{figure}

As the next example we will use the structure of a surface subsystem topological CSS code introduced in \cite{Bravyi12}.

\textit{Code geometric structure.} Fig. \ref{bravyi}(a) presents planar version of the code. Qubits are placed on the vertices of the lattice. Stabilizers are generated by pairs of 3-weight operators which are placed on the opposite side of every square frame and by 2-weight operators on the boundary. For $L$ denoting the number of columns of the code structure, there are $L^{2}+1$ logical qubits encoded. Bottom triangles of every square frame are logical operators of $L^2$ logical qubits, and, by measuring them, we can make the measurement of 6-weight stabilizers local on the price of destroying $L^2$ logical qubits, so that only one logical qubit is left for the encoding purposes. Its $Z_{L}$ ($X_{L}$) operator is an arbitrary string of $Z$ ($X$) operators joining bottom (left) boundary with its upper (right) counterpart. W.l.o.g. we consider $Z_{L}$ operator along left-most column and $X_{L}$ operator along bottom line. Their intersection point defines the vertex to host a physical qubit we want to encode. As usually, we prepare qubits in $|0\rangle$ ($|+\rangle$) states along $Z_{L}$ ($X_{L}$) operators. Additionally, we divide a lattice into two parts along a diagonal and surround the lines of qubits in $|0\rangle$ and  $|+\rangle$ states with qubits in $|0\rangle$ and $|+\rangle$ states, respectively (see Fig. \ref{bravyi}(b)).

\textit{Encoding procedure.} All $X$-type and $Z$-type syndrome measurements are performed and then corrections are applied. Perfect preparation of $|0\rangle$ and $|+\rangle$ states ensures that the parity detected by logical operators depends only on the state of a physical qubit to be encoded as long as all syndromes are corrected by chains of $X$ and $Z$ Pauli operators toward right and upper boundary, respectively, so that logical operators $Z_{L}$ and $X_{L}$ remain untouched. Because all stabilizers in the code are represented by plaquettes (except for boundary stabilizers) we consider a graphical representation of the recovery procedure in a virtual lattice, where virtual edges are associated with code qubits and virtual vertices with plaquettes. Due to the structure of the code, we have to create separate virtual lattices for correcting bit and phase errors.  Let us focus on correction of $X$-type errors. The scheme of single shot encoding in the virtual lattice is shown in the Fig. \ref{bravyi_virtual}(a). Thin lines represent the original lattice. Each diamond corresponds to $6$-weight $Z$-type stabilizer and each circle to $2$-weight $Z$-type boundary stabilizer. Virtual links marked in red, blue and green represent qubits prepared in $|\Psi\rangle$, $|+\rangle$ and $|0\rangle$ states, respectively. After measuring $6$-weight and $2$-weight $Z$-stabilizers, we record measurement results on virtual nodes. Then we join the defects together or with one of the boundaries without touching a line of logical operator. Procedure for repairing phase errors goes similarly but virtual lattice is rotated by $90$ degrees, see Fig. \ref{bravyi_virtual}(b).

\textit{Decoding procedure.} All qubits situated in the line of logical operator $X_{L}$ ($Z_{L}$) except for the red one are measured in $|+\rangle$, $|-\rangle$ ($|0\rangle$, $|1\rangle$) basis. If parity obtained from the results is odd, phase (bit)-flip has to be applied to the red qubit. \\

\subsection{Haah code}
\begin{figure}
\includegraphics[scale=0.5]{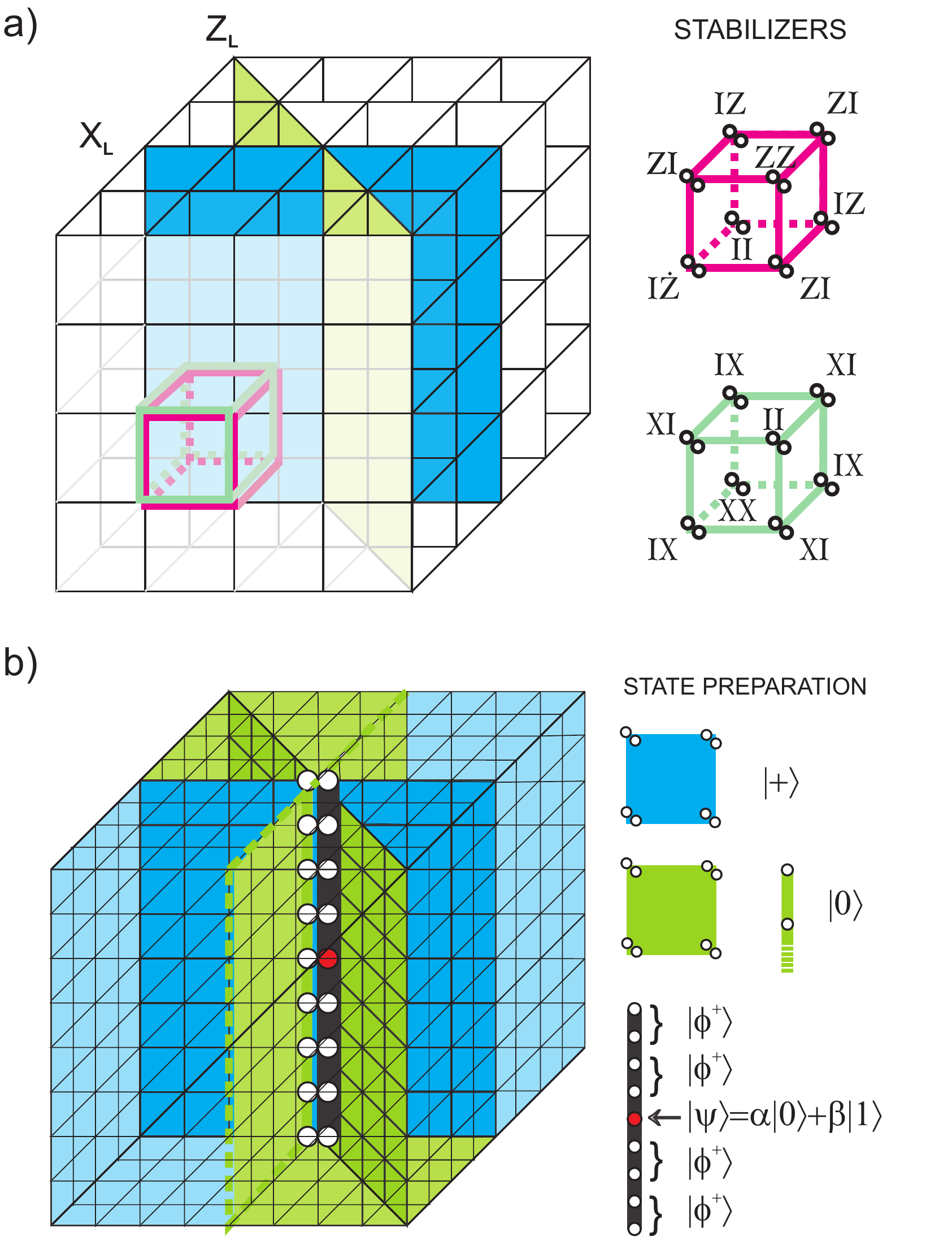}
\caption{\label{haah}
\textbf{(a) Haah code.} Each vertex is associated with two code qubits. Stabilizer generators of types $X$ and $Z$ form cubes. Exemplary logical operators $X_{L}$ and $Z_{L}$ associated with blue and green planes cross on a line. \textbf{(b) Preparation of a lattice.} Logical operators $X_{L}$ and $Z_{L}$ (blue and green plane) cross at a line. Due to the structure of the code, there are two code qubits placed on each vertex in that line. First qubit of each pair is initialised in state $|0\rangle$. In the centre of a thick black line composed of second qubits of every pair, red qubit is inserted. Remaining qubits lying on that line are combined in pairs and every such pair is prepared in  maximally entangled state $|\phi^{+}\rangle$. Qubits in green and blue regions are initialised in states $|0\rangle$ and $|+\rangle$, respectively.}
\end{figure}

All topological CSS codes considered above fulfil the requirement that logical operators $X_{L}$ and $Z_{L}$ cross only at one point. Hence, the proposed encoding/decoding scheme could be there directly implemented. Below we discuss how to modify the procedure in order to encode an unknown state into topological CSS codes where both logical operators act nontrivially on larger (odd) number of qubits. We will investigate the case of three dimensional topological stabilizer CSS code of \cite{Haah11}.

\textit{Code geometric structure.} The code is constructed on a $\mathbb{Z}^{3}_{L}$ lattice with 2 physical qubits on its every vertex. Stabilizers act on every cube of the lattice and are generated by operators depicted in Fig. \ref{haah}(a). We denote by $L$ the size of the lattice. For odd $L$ in the range $2\leq L\leq200$ and $L\neq 15n$, $L\neq 63n$, where $n\in\mathbb{N}$, there are two logical qubits with logical operators acting on planes of physical qubits in the lattice. We show here the procedure for one of these qubits. Logical operators are given by $Z_{L}=\underset{(1,-1,0)-plane}{\otimes}ZZ$ and $X_{L}=\underset{(1,0,0)-plane}{\otimes} IX$, where planes of physical qubits are identified by vectors orthogonal to them. As logical operators cross on a line, the requirement that they intersect on a single point is not fulfiled. 
Below we show how to prepare a lattice in order to still be able to encode a qubit in unknown state into this code. Let's split a line of intersection between $X_{L}$ and $Z_{L}$ into $2$ strings, first composed of first qubits from each pair lying on that line, and second composed of qubits on which both $X_{L}$ and $Z_{L}$ act nontrivially, i.e. second qubits from each pair. We substitute one qubit from second string by a physical qubit in a state $|\Psi\rangle$ we want to encode. The choice of its position on the string is such that it divides the odd number of qubits into regions consisted of even number of qubits. Inside these regions, qubits are grouped into pairs. Each pair is {\it locally} prepared in a singlet state $|\phi^{+}\rangle=\frac{1}{\sqrt{2}}|00\rangle+|11\rangle$  (see Fig. \ref{haah}(b)). The rest of the protocol is similar to the original procedure: qubits on which only operator $X_{L}$ ($Z_{L}$) acts nontrivially, are initialised in states $|+\rangle$ ($|0\rangle$). Remaining qubits are prepared in such a way that areas of $|+\rangle$ and $|0\rangle$ states surround the planes of logical operators from one or both sides (blue and green regions in Fig. \ref{haah}(b)).

\textit{Encoding procedure.} Stabilizers depicted in Fig. \ref{haah}(a) are measured and then corrections are applied. During the correction stage, we have to count the number of times that anticommuting operators $X$ ($Z$) touch the plane of logical operator $Z_L$ ($X_L$). Another logical operator $Z_L$ ($X_L$) is applied to the code whenever this number is odd. It is easy to show that the above scheme is a proper encoding procedure due to the stabilizer measurements and subsequent corrections. The following applications of logical operators do not take us out of the codespace. Preparation of singlet states, stabilised by pairs of $XX$ and $ZZ$ operators, as well as preparation of $|+\rangle$ and $|0\rangle$ states on given positions, ensures that the parity measured by $Z_{L}$ and $X_{L}$ operators depends only on the physical state $|\Psi\rangle$ placed on the intersection line. As all stabilizers commute with logical operators, we cannot affect parity by stabilizer measurements. Applying proper logical operators triggered by the odd number of crossings of the logical operator by anticommuting correction chains ensures that the logical qubit of the code is in the $|\Psi\rangle_L$ state.

\textit{Decoding procedure.} The crucial thing about decoding is to establish a parity on areas where logical operators are defined with exception of a place occupied by the red qubit. In previous cases, we performed single qubit measurements in $X$, $Z$ eigenbasis within mentioned areas affected by $X_{L}$, $Z_{L}$, respectively. These anticommuting measurements were allowed since they were made on separate qubits (the only qubit which was under the influence of both logical operators was not measured). 
Here, as logical operators intersect on the line, and solely the one qubit lying on it is not measured, we need to modify our procedure in order to avoid performing noncommuting measurements on the rest of the qubits contained within that line. Instead of measuring single qubit operators we measure operators $X_iX_{i+1}$ , $Z_iZ_{i+1}$ on neighboring qubits above and below red qubit. The remaining procedure goes as before, namely in places where either $X_{L}$ or $Z_{L}$ operator acts nontrivially, individual qubits are measured in $X$ or $Z$ eigenbasis, then the parity of truncated logical operators is computed. When the obtained parity is odd, phase or bit-flip has to be applied to the red qubit. \\

\section{Encoding/decoding procedure in a presence of noise}
\label{sec:noise}
\begin{figure}
\centering
\includegraphics[width=85mm]{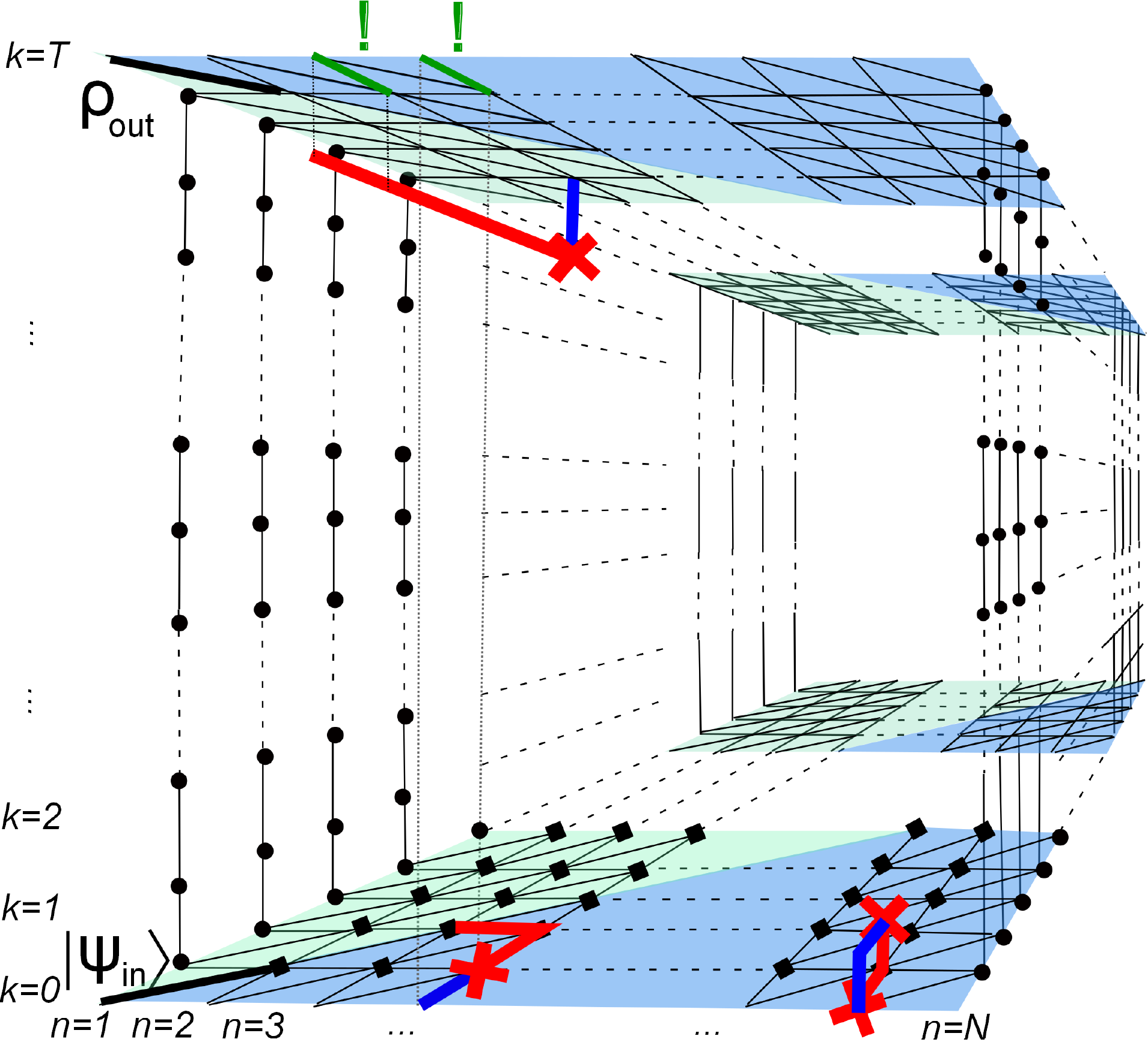}
\caption{\label{bravyi_3D}
\textbf{Graphical representation of encoding, storage and decoding procedure in subsystem surface code}. In the procedure Z-type errors happen with probability $p$, likewise errors on $X_s$ stabilizer measurements performed between every horizontal plane. Horizontal links denote code qubits, vertical links are places to store the outcomes of syndrome measurements. Vertices in the lattice marked by red crosses correspond to nontrivial syndrome in the first horizontal slice $k=0$ and changes of it in all subsequent $k=1,2,....T$ slices. Both, top and bottom planes are divided into $3$ parts: black, blue and green. For the latter one black qubit is prepared in $|\Psi_{in}\rangle$ state, and blue (green) qubits in $|+\rangle$ ($|0\rangle$) states. The top plane illustrate decoding stage where syndromes are computed from single qubit measurements made in $X$ eigenbasis. Three examples of actual error chains (elements of E) are represented by red thick lines. Each ends with a defect/defects (red crosses) or on a boundary (front wall or green regions). Corrections ($E_{min}$) are performed along the shortest paths due to a given metric and are represented by blue thick lines. Paths from $E+E_{min}$ are either nontrivial or trivial chains, two examples of former and one of latter are shown in the picture. Nontrivial paths from $E+E_{min}$ are \textit{dangerous} since they change a parity of $X_L$ operator (see exclamation marks) and logical error may occur. }
\end{figure}

In this section, we consider a noisy scenario, where qubits are subjected to bit-flip and phase-flip errors (while being stored and prepared) and where measurements can be faulty. Such noisy syndrome measurement is modelled by flipping the ideal measurement outcome with some probability.  We assume that probabilities of a bit-flip, phase-flip and syndrome measurement errors are equal to $p$.
The general idea is to prepare all qubits as described in Section III, measure $X_s$ and $Z_p$ stabilizers many times in the area confined by the whole lattice (except for the last time step where $X$ and $Z$ operators are measured), store all error syndromes and use them to apply error correcting procedure. Below we describe it in detail for Bravyi subsystem topological code.
One can perform similar analysis for Kitaev toric code and defected lattice code. However, we have not been able to 
find a noisy scheme for Haah code.

Since the schemes for protecting quantum information from bit-flip and phase-flip errors are considered in the separate virtual lattices (Fig. \ref{bravyi_virtual}(a) and Fig. \ref{bravyi_virtual}(b)), we focus on phase-flip errors. The reasoning for bit-flip errors is similar.
Our encoding/storage/decoding scheme which protects against phase-flip errors can be graphically represented on the $3D$ lattice. Each horizontal plane corresponds to a lattice of qubits at consecutive time steps $k=0,1,2...T$, see Fig. \ref{bravyi_3D}. Syndrome measurements are performed on qubits placed on each horizontal plane, and their outcomes are stored in vertical links. Bottom plane is prepared identically to the $2D$ virtual lattice considered in the case of  noiseless scenario (Fig. \ref{bravyi_virtual}(b)), i.e. there are two separate regions of qubits initialised in $|+\rangle$ (marked in blue) and $|0\rangle$ (marked in green) states. A qubit in a $|\Psi_{in}\rangle$ state which we want to encode is chosen to be placed in the front left corner (black thick link in Fig. \ref{bravyi_3D}). The top plane ($k=T$) is graphically divided into three parts: black, blue and green. 

After preparation of a lattice, we measure $X_s$ stabilizers in all time steps except the last one (we also measure $Z_p$ stabilizers but they are irrelevant when considering phase-flip errors). Squares and circles in Fig. \ref{bravyi_3D} (see also Fig. \ref{bravyi_virtual}(b)) represent $6$-weight and $2$-weight $X_s$ stabilizers, respectively. By red crosses we denote defects, i.e. nontrivial error syndromes in the first time step and changes of syndromes in all subsequent time steps. In the last time step $T$ we measure $X$ operator on each qubit in blue triangle (which corresponds to the bottom triangle where $|+\rangle$ states were prepared) and $Z$ operator on each qubit in green triangle (which corresponds to the bottom triangle where $|0\rangle$ states were prepared).     
Subsequently, a syndrome $X_s$ is computed by classically multiplying the outcomes of single qubit measurements  for vertices in top triangle marked in blue. Hence, for blue region we obtain ideal syndrome. For green region the syndrome is unknown. The consequences of that will be discussed later on.

The next step is to apply corrections. We use the following terminology. $S$ indicates a set of links with nontrivial error syndrome $-1$. Its boundary $\delta S$ is determined by defects (red crosses), apart from the green region on the top slice. Links on which an actual error occurs compose a set of errors $E$ (horizontal and vertical red thick lines in Fig. \ref{bravyi_3D}). The boundary of $E$ is associated with vertices that lie outside of top green region and on which syndrome changes or, in the case of $k=0$ plane, is nontrivial. Namely, $\delta E=\delta S$, apart from the top green region. Clearly, there exist many hypothetical sets of errors $E'$ that could lead to the same defects ($\delta S$). The most probable set (i.e. containing the shortest paths of errors due to a metric presented below) is denoted by $E_{min}$, see horizontal and vertical blue thick lines in Fig. \ref{bravyi_3D}. Weights of links building a $3D$ lattice are given by a formula $-\log \frac{p_i}{1-p_i}$ \cite{kitajewpreskill-ftmem}, where $p_{i}$ stands for the probability of error on the $i$-th qubit. For qubits in the green bottom triangle $p_{i}=\frac{1}{2}$, as they are prepared in $|0\>$ state, i.e. they are not prepared in eigenstate of $X$-type operators. Since within the green top triangle single-qubit measurements are performed in $\{|0\rangle, |1\rangle \}$) basis, also for that area we assign $p_{i}=\frac{1}{2}$. Black qubit is exposed to storage error, thus $p_{i}=p$. For qubits in the blue top triangle $p_{i}=p$ since they are subjected to preparation error. To qubits located in the blue bottom triangle we ascribe $p_{i}=p$ as single-qubit measurement gives erroneous outcome with probability $p$. For qubits in all other time slices $p_{i}=p$ due to storage error. Probability $p_{i}=p$ is also assigned to all vertical links as syndrome measurement gives there erroneous outcome with probability $p$. Given those weights we can determine $E_{min}$ by minimising the sum of weights. If the set of actual errors $E$ is the most probable one, then $E_{min}=E$. 

To recover from Z-type errors we have to connect observed defects (by chains of Pauli operators $Z$) with each other or with one of the code boundaries. We recognise code boundaries as front and back vertical planes (geometrical boundaries) as well as green triangles (due to ascribed weights equal to $0$ within these regions). In order to maximise a probability of successful correction, we annihilate defects along the shortest paths from the set $E_{min}$. Hence, a disjoint union $S+E_{min}$ and also $E+E_{min}$ have no boundary. Therefore, within $E+E_{min}$ there are only closed or nontrivial paths.

In order to proceed to the decoding procedure, after determining the set $E_{min}$, we need to calculate a parity of a line belonging to the logical operator $X_L$, excluding a black qubit at position $n=1$. We pretend that the true error that occurred during storage process is given by $E_{min}$. Then the mentioned parity could have been modified by applying noncommuting corrections ($Z$ operators) from the set $E_{min}$. To see if the parity was affected we make a projection of $E_{min}$ on the $k=T$ plane and count the number of times it crosses a curved line of logical operator. If this number is odd, parity was affected and has to be flipped. Afterwards, when the corrected parity of truncated logical operator $X_T$ is odd, we apply phase flip to the black qubit, obtaining $\rho_{out}$. Thus, after protecting the state for time $T$, it is possible to decode it.

\begin{figure}
\includegraphics[scale=0.5]{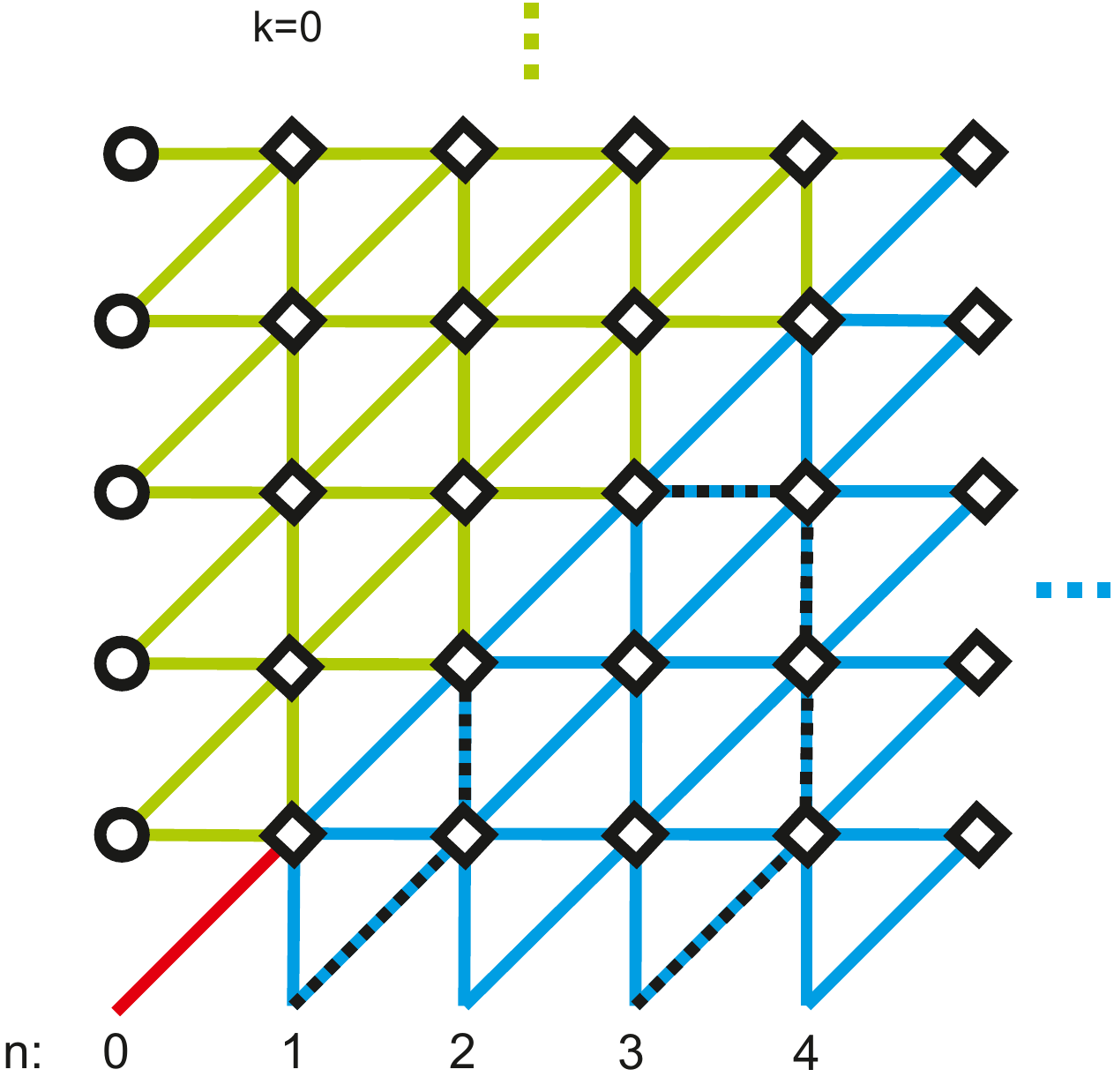}
\caption{\label{bravyi_phase}
\textbf{Calculation of lower bound on the protocol fidelity for encoding/decoding phase.} Starting points for paths of errors and corrections which affect a line of logical operator $X_L$ are labelled on each horizontal plane (here only a case for $k=0$ is shown) by index $n=0,1,...,\frac{N-1}{2}$, where $N$ is a distance of a code. Examples of such paths denoted by dashed lines are depicted for $n=1;3$ and $k=0$. Green area is considered as a boundary so the lengths of paths running within that region are equal to $0$.}
\end{figure}

If the true errors were just $E_{min}$, we would decode the initial state perfectly (up to the last flip, which is noisy). Even more: if 
$E+E_{min}$ had an even number of nontrivial paths, we still would have such perfect decoding. Thus the recovery procedure is not successful when, as a result of applied corrections, we create an odd number of nontrivial paths in $E+E_{min}$ set. This is due to the fact that such paths cross a logical operator $X_L$ an odd number of times, i.e. they change its parity. The probability of an error is bounded by the probability that the set $E+E_{min}$ has a nontrivial path. In order to obtain upper bound on probability of erroneous encoding, storage and decoding of phase we overestimate the number of nontrivial paths of length $l$, multiply it by probability that a particular path of length $l$ is in $E+E_{min}$, and sum over $l$.
Let us choose the line of logical operator $X_L$ to be a curved front line in a virtual lattice (denoted in Fig. \ref{bravyi_virtual}(b) by thick black line for $k=0$). Each nontrivial path or its projection on a plane $k=T$ has to cross the logical operator in the slice $k=T$. Therefore, we take into account every path that starts at any point $(n,k)$ on the front geometrical boundary and goes toward the opposite boundary, or top or bottom green regions (which are boundaries as well).  Indices $n=0,1,2,...,\frac{N-1}{2}$ and $k=0,1,...,T$ enumerate points in space and time directions respectively. Fig. \ref{bravyi_phase} shows bottom surface of the $3D$ code structure for $k=0$. An odd number $N$ denotes a distance of a code. Now we divide paths into two sets. The first set consists of nontrivial paths of length $l=1$. There are $4$ such paths which start at points $(n,k)=(0,0),(1,0),(0,T),(1,T)$ and each path occurs with probability $p$. The second set consists of nontrivial paths of length $l \geq 2$.  At any point in front geometrical boundary can start at most  $8^l$ nontrivial paths of length $l$ which are in the set $E+E_{min}$. The factor $8^l$ comes from fact that once a path reaches a particular point it can go to no more than $8$ other points. We note that not all lengths are allowed. More precisely at point $(n,k)$ there can only start nontrivial path of length $l\geq \min(k+n,(T-k)+n)$. The probability that such a path is in $E+E_{min}$ is $prob(l)\leq(2\sqrt{p(1-p)})^{l}$ \cite{kitajewpreskill-ftmem}.
Taking it all into account we can write the probability of failure as
\begin{widetext}
\begin{multline}\label{p2}
P_{fail}\leq\ 4p + (2\sum_{k=1}^{T-1}\sum_{l=\min(1+k,1+(T-k),N)}^{\infty}+\sum_{n=1}^{1}\sum_{k=0}^{0}\sum_{l=\min(n+1+k,n+1+(T-k),N)}^{\infty}+\sum_{n=1}^{1}\sum_{k=T}^{T}\sum_{l=\min(n+1+k,n+1+(T-k),N)}^{\infty}+ \\ +\sum_{n=2}^{\frac{N-1}{2}}\sum_{k=0}^{T}\sum_{l=\min(n+k,n+(T-k),N)}^{\infty})\alpha^{l}= \\ = 4p + (2\sum_{k=1}^{T-1}\sum_{l=\min(1+k,1+(T-k),N)}^{\infty}+2\sum_{l=\min(2,N)}^{\infty}+\sum_{n=2}^{\frac{N-1}{2}}\sum_{k=0}^{T}\sum_{l=\min(n+k,n+(T-k),N)}^{\infty})\alpha^{l} \\
\end{multline}
\end{widetext}
where $\alpha=16\sqrt{p(1-p)}$ and $\alpha\leq1$ for $p\leq 0.0039$. \\
In the limit of large code size $N\rightarrow \infty$ we obtain
\be\label{bound3}
\lim_{N\rightarrow \infty} P_{fail}\leq 4p+\frac{2(2-\alpha)^2\alpha^2}{(1-\alpha)^3}.
\ee
RHS of (\ref{bound3}) was largely overestimated and decreases below $1/2$ for $p\leq 0.000154$.
The bound for a case of bit encoding/decoding is slightly different:
\be\label{bound4}
\lim_{N\rightarrow \infty} P_{fail}\leq 10p+\frac{2(3-\alpha)(2-\alpha)\alpha^2}{(1-\alpha)^3}.
\ee
RHS of (\ref{bound4}) was largely overestimated and decreases below $1/2$ for $p\leq0.0001085$. 
Hence the fidelity of the protected qubit is $1-\mathcal{O}(p)$ for small $p$.

\section{Conclusions}
In this paper we introduced a simple, single shot procedures for encoding/decoding an unknown state into/from logical subspace of CSS codes. The encoding procedure relies on preparing a system in a way that makes the parity of logical operators dependent only on the state of a selected qubit of the system, and on driving the state of the system into a logical subspace by sequence of operations that commute with logical operators. 

For topological subsystem code \cite{Bravyi12} we calculated lower bound on the fidelity of a process of encoding/storage/decoding of an unknown state under the assumption of presence of preparation and syndrome measurement error as well as local noise introducing phase and bit storage errors (we assumed that probability of each error is $p$). Noisy syndrome measurement was modelled by ideal measurement followed by classical error acting on the outcome with probability $p$. In the limit of large code size, the bound for the fidelity was found to be $1-\mathcal{O}(p)$. In a similar way 
one can obtain fidelity bounds for Kitaev code on a torus and planar code with holes.

In our analysis, we have assumed, that there is no back reaction from syndrome measurements to the code.
For Kitaev planar surface code, Fowler et al. \cite{fowler10} proposed modified measurement schemes aimed at avoiding the effect of back-action. The analysis, affected the value of the threshold, but not its existence. 
Similarly, we believe that the implementation of these ideas could lead to modification of the bound for fidelity of encoding/storing/decoding process obtained in this paper. However its existence should be not affected.  

Proposed general encoding/decoding processes require entanglement preparation/measurement, hence it may be, in principle, nonlocal for some codes. However, when the qubits at which logical operators cross are situated on the adjacent vertices of the code structure, this can be achieved locally, as in the Haah code.  For the latter code we have provided noiseless version, and we are leaving as an open question, how to encode unknown state into Haah code in presence of noise.

\section*{Acknowledgements}
This work was supported by ERC Advanced Grant QOLAPS  and National Science
Centre project Maestro DEC-2011/02/A/ST2/00305. MH and PM acknowledge support of MNiSW Ideas-Plus Grant IdP2011000361. PM was supported by the International PhD Project "Physics of future quantum-based information technologies": grant MPD/2009-3/4 from Foundation for Polish Science.

\end{document}